\documentclass[aps,showpacs,twocolumn,floatfix,pra]{revtex4}
\usepackage{graphicx}
\usepackage{amsfonts}
\usepackage{amsmath}
\usepackage{amssymb}

\newcommand{\beq}{\begin{equation}}
\newcommand{\eeq}{\end{equation}}
\newcommand{\beqn}{\begin{eqnarray}}
\newcommand{\eeqn}{\end{eqnarray}}
\newcommand{\bearr}{\begin{array}}
\newcommand{\enarr}{\end{array}}

\begin{document}


\title{Periodically driven  Quantum Ratchets: Symmetries and Resonances}

\author{S. Denisov$^{1}$, L. Morales-Molina$^{2}$, S. Flach$^{2}$, and P. H\"{a}nggi$^{1}$}
\affiliation{$^1$ Institut f\"ur Physik, Universit\"at  Augsburg,
       Universit\"atsstr.1, D-86135 Augsburg, Germany}

\affiliation{$^2$ Max-Planck-Institut f\"ur Physik Komplexer
Systeme, N\"othnitzer Str. 38, 01187 Dresden, Germany}
\date{\today}
\vskip 2.cm
\begin{abstract}
We study the quantum version of a tilting and flashing Hamiltonian
ratchets, consisting of a periodic potential and a time-periodic
driving field. The system dynamics is governed by a Floquet
evolution matrix bearing the symmetry of the corresponding
Hamiltonian. The dc-current appears due to the desymmetrization of
Floquet eigenstates, which become transporting when all the relevant
symmetries are violated. Those eigenstates which mostly contribute
 to a directed transport reside in phase space regions
corresponding to classical resonances. Quantum dynamics leads to the
dependence of the average velocity on the initial phase of the
ac-field. A resonant enhancement (or suppression) of the dc-current,
due to avoided crossings between different Floquet states takes
place upon tuning some
 control parameters. Our studies are predominantly aimed at
experimental realizations of ac-driven quantum ratchets with cold
atoms.
\end{abstract}
\pacs{05.45.Mt, 05.60.-k, 32.80.Pj}

\maketitle

\section{Introduction}

Ratchets are viewed as realizations of systems which produce a
directed current from a fluctuating environment in the absence of
gradients and net forces \cite{first, sec, Reim, Astum,Hanggi_last}.
Initially proposed as an abstract physical model
 for the understanding of a micromolecular machinery \cite{sec},
 ratchet systems have found diverse applications in
many areas \cite{Reim, Astum, Hanggi_last}, from mechanical devices
\cite{Slava} up to the quantum systems \cite{Qu1, Qua2, Qu3, Qu4,
jong}. Among the areas with growing interest is the study of ratchet
dynamics for cold atoms \cite{cold}. Hamiltonian ratchets
\cite{Flach1, Flach2, Den, Ketz, Goy} with the corresponding
symmetry predictions \cite{Flach1} have recently been successfully
realized with cold Rubidium and Cesium atoms in optical lattices
with a two-harmonics driving and a tunable weak dissipation
\cite{ren1}. In these experiments, the mechanism of the Sisyphus
cooling \cite{Sis} has been used in order to furnish initial
conditions in form of an \textit{optical lattice}: an ensemble of
atoms localized in the wells of a periodic potential created by
laser beams. In the momentum space this corresponds to a narrow
distribution near the momentum $p=0$. This is essential for the
observation of the rectification effect, since the asymptotic
current tends to zero for broad initial distributions in momentum
 space.

To describe the ratchet dynamics of the thermal cloud of cold atoms
in optical lattices \cite{ren1}, a classical model has been used
\cite{Flach1, Flach2, Den}. However, as the cloud of atoms in a far
detuned optical potential \cite{rew} is cooled further down, one can
obtain a system where quantum effects become relevant.
 Many
studies of quantum ratchets are based on the kicked rotor model
\cite{Ketz, Qu1, Qu1_2, Qu2}, which is readily treated numerically.
In reality kicks  may heat the system and degrade quantum effects.
It is perhaps for that reason, that other successful experiments
\cite{ren1} use a time continuous drive, which is also the choice in
the present work.

 The  paper is organized as follows.
In Section II we introduce the model for the quantum tilting ratchet
and discuss its properties. The dynamics of the system  is studied within the Floquet theory.
We use the Husimi representation in order to link
eigenstates to manifolds in the classical phase space. In section III
we study the relevant symmetries, whose violation leads to the
directed current, and the desymmetrization mechanisms.
 Section IV is
devoted to the dynamics of current rectification. We begin by discussing
the dependence of the current on the initial condition. Then we
investigate the appearance and controlling of quantum resonances
that lead to a significant enhancement of directed transport. In
Section V we introduce the realization of a quantum flashing ratchet system.
 Finally, in Section VI we summarize our
results. Some of our important technical details are deferred
 to the Appendices A, B and C.

\section{model set-up}
Let us start with a cloud of atoms, placed into a periodic potential
(formed by an optical standing wave) and exposed to an external ac
field. Assuming a low density, we neglect the interactions among the
atoms. Hence, our problem can be described by
the Schr\"{o}dinger equation

\begin{equation}
i \hbar \frac{\partial}{\partial t} |\psi(t)\rangle =
H(t,t_{0})|\psi(t)\rangle, \label{eq:Schrodin}
\end{equation}
where the  Hamiltonian $H$ is of the form \cite{ren1}
\begin{equation}
H(x,p,t)=\frac{p^{2}}{2}+(1+\cos(x))-x E(t-t_{0}).
\label{eq:ham}
\end{equation}
Here $E(t)$ is an external periodic field of zero mean,
$E(t-t_{0}+T)=E(t-t_{0})$, $\int_0^T E(t) dt=0$, and $t_{0}$ is the
initial time.

The Hamiltonian (\ref{eq:ham}) is periodic in time with period $T$.
The solutions  $|\psi_{\alpha}(t)\rangle =
U(t,t_{0})|\psi_{\alpha}(0)\rangle$ of the Schr\"{o}dinger equation
(1) can be characterized by the eigenfunctions of $U(t_0+T,t_0)$
which satisfy the Floquet theorem: $ |\psi_{\alpha}(t)\rangle=
e^{-i\frac{E_{\alpha}}{T}t} |\phi_{\alpha}(t)\rangle$,
$|\phi_{\alpha}(t+T)\rangle=|\phi_{\alpha}(t)\rangle$. The
quasienergies $E_{\alpha}$ $(-\pi < E_{\alpha} < \pi)$ and the
Floquet eigenstates can be obtained as solutions of the eigenvalue
problem of the Floquet operator
\begin{equation}
U(T,t_{0})|\phi_{\alpha}(t_{0})\rangle = e^{-i
E_{\alpha}}|\phi_{\alpha}(t_{0})\rangle. \label{eq:Floquet}
\end{equation}
The Floquet eigenstates provide a complete orthonormal basis and the
stroboscopic quantum state can be expressed as \cite{Grif}
\begin{equation}
|\psi(mT, t_{0})\rangle = \sum_{\alpha} C_{\alpha}(t_{0}) e^{-im
E_{\alpha}}|\phi_{\alpha}(t_{0})\rangle, \label{eq:expans}
\end{equation}
where the coefficients $\{C_{\alpha}\}$ depend on $t_0$. For later
convenience the integer $\alpha=0,1,2..$, sorts the states
$|\phi_{\alpha}\rangle$  such, that the mean kinetic energy $\langle
p^{2}\rangle_{\alpha} \equiv 1/T \int_0^T \langle \phi_{\alpha} |
\hat{p}^2 | \phi_{\alpha} \rangle dt_0 $ monotonically increases.

By using the gauge transformation, $|\psi \rangle \rightarrow
\exp(-\frac{i}{\hbar}x\int_{0}^{t}E(t')dt') |\psi \rangle$
\cite{gauge1}, we transform the original Hamiltonian
(\ref{eq:ham}) into a spatially periodic one (see Appendix A). Then
the solution of the time-dependent Schr\"{o}dinger equation for the
new Hamiltonian may be written as
\begin{equation}
|\psi(t)\rangle =e^{-\frac{i}{\hbar} \int_{0}^{t} \{\frac{1}{2}[\hat{p}-
A(t',t_{0})]^{2}+(1+\cos x)\}dt'} |\psi(0)\rangle,
\label{eq:evolution}
\end{equation}
with the vector potential  $A(t)=-\int_{0}^{t}E(t')dt'$. Due to
discrete translational invariance and Bloch's theorem all Floquet
states are characterized by a quasimomentum $\kappa$ with
$|\phi_{\alpha}(x+2\pi)\rangle = {\rm e}^{i \hbar \kappa}
|\phi_{\alpha}(x)\rangle$.

Here we choose $\kappa=0$ which corresponds to initial states
where atoms equally populate all (or many) wells of the spatial
potential. The
wave function is expanded in the plane wave
eigenbasis of the momentum operator
$\hat{p}$, $|n \rangle=\frac{1}{\sqrt{2 \pi}} e^{i n x}$, viz.
\begin{equation}\label{Eq:wave}
|\psi(t)\rangle=\sum_{n=-N}^{N} c_{n}(t) |n \rangle.
\end{equation}

A detailed discussion of the two numerical procedures used to
integrate Eq.(\ref{eq:Schrodin}) is presented in the Appendix
\ref{nume}.

To examine the morphology of the quasienergy states we use the
Husimi representation \cite{Husimi}
\begin{equation}\label{eqhusi}
\rho(\langle  x \rangle,\langle p \rangle ) =\frac{1}{2\pi}|\langle
\psi| \Phi_{\langle  x \rangle,\langle p \rangle}  \rangle|^2,
\end{equation}
where
\begin{equation}
\Phi_{\langle x \rangle ,\langle  p \rangle}(x)=\frac{1}{(2\pi
  \sigma^2)^{1/4}}\exp\left\{-\frac{(\langle x \rangle-x)^2}{4\sigma^2}+i
  \langle p \rangle x \right\}
\end{equation}
 with $\sigma=(\hbar/2)^{1/2}$. Here
$\langle x \rangle$ and $\langle p \rangle$ stand for the average of
the position and momentum respectively. The Husimi representation
provides an  insightful, coarse grained description in the
phase-space \cite{Husimi}.

\section{Relevant symmetries and their violations}

\subsection{Classical limit}

Let us briefly outline the classical case. In this limit the system
is generically characterized by a nonintegrable dynamics with a
mixed phase space containing both chaotic and regular areas in the
three-dimensional phase space $(x,p,t)$ \cite{Ham_cl}. Due to time
and space periodicity of the classical equation of motion $\ddot
x=-\frac{\partial
 H}{\partial x}$, we can map the
original three-dimensional phase space  onto a two-dimensional
cylinder, $\texttt{T}^{2}=(x \; mod 1,p)$, by using  the
stroboscopic Poincar$\acute{e}$ section after each period
$T=2\pi/\omega$, cf. Fig.1(a). This provides a helpful visualization
of the mixed phase space structure. A stochastic layer, located near
the line $p=0$, originates from the destroyed separatrix of the
undriven system, $E(t)=0$.  The chaotic layer is confined by
transporting $KAM-tori$, which originate from perturbed
trajectories of particles with large kinetic energies, $\langle
p^{2} \rangle$. The stochastic layer is not uniform and contains
different regular invariant manifolds - regular islands, unstable
periodic orbits and cantori \cite{Ham_cl}.

There are two symmetries which need to be broken to fulfill the
necessary conditions for the appearance of a dc-current
$J=\lim_{t\rightarrow\infty} \frac{x}{t}=\langle \dot{x }\rangle$
\cite{Flach1}. If $E(t)$ is shift symmetric, i. e. $E(t)=-E(t+T/2)$,
the symmetry
\begin{equation}
S_{a}: (x, p, t) \rightarrow (-x, -p, t+T/2), \label{eq:Sa}
\end{equation}
is realized. Furthermore, if $E(t)$ is symmetric, i. e.
$E(t+t_{s})=E(-t+t_{s})$ at some appropriate point $t_{s}$, the
symmetry
\begin{equation}
S_{b}: (x, p, t) \rightarrow (x, -p, -t+ 2t_{s}), \label{eq:Sb}
\end{equation}
is holds. Any trajectory, when transformed using one of the
abovementioned symmetry operations, yields again a trajectory
of the system but with opposite velocity. Assuming that ergodicity
holds in the stochastic layer (which means that all its average
characteristics are independent on the initial conditions, and that symmetry-related trajectories
have the same statistical weight, we conclude that the asymptotic
velocity within the chaotic layer is zero.  So, whenever $S_{a}$
and/or $S_{b}$ are realized, directed transport is forbidden inside the
chaotic layer
\cite{Flach1}.
\begin{center}
\begin{figure}[t]
\begin{tabular}{cc}
\includegraphics[width=3.5cm,height=4cm]{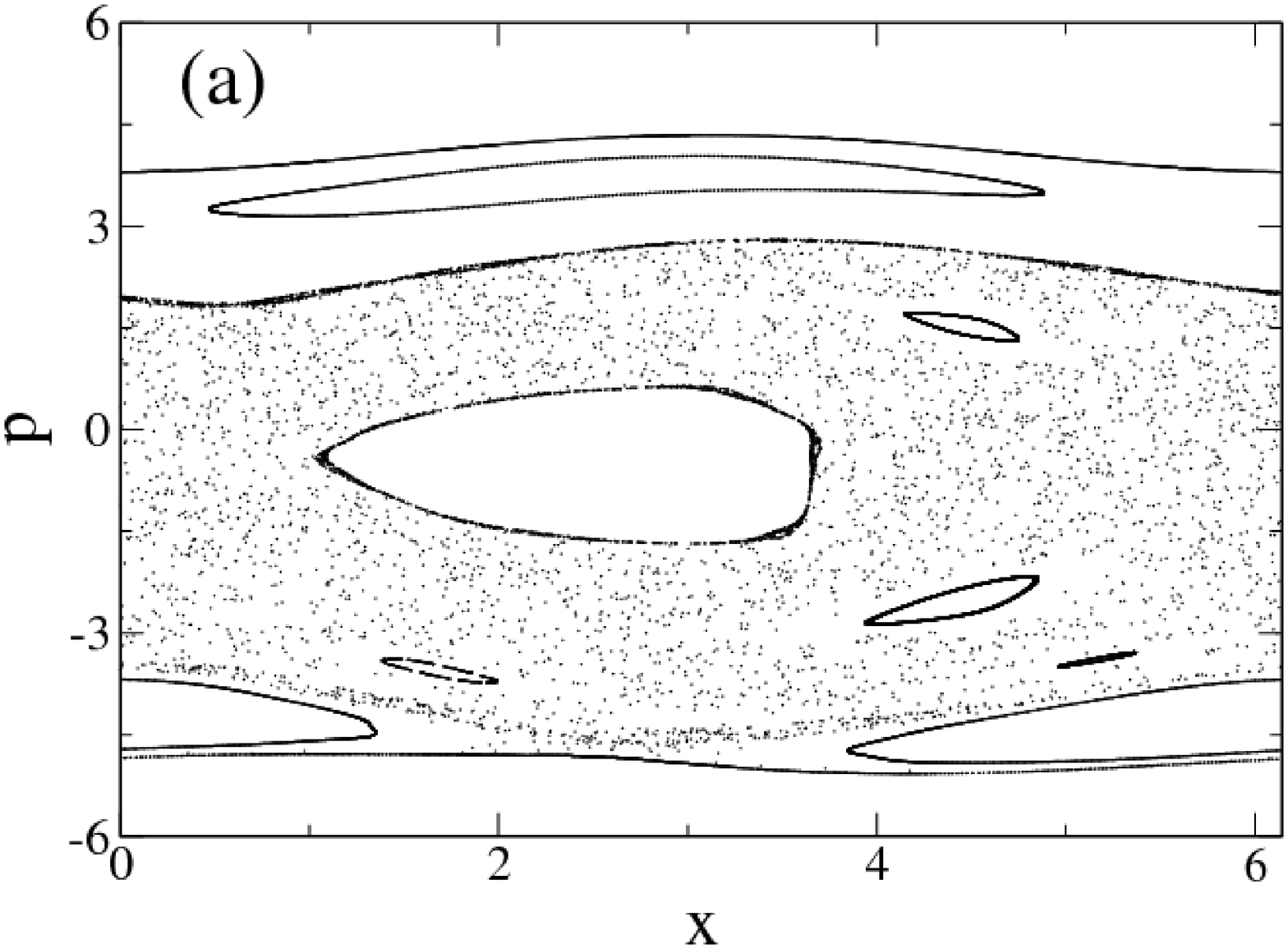}
\includegraphics[width=4.2cm,height=4cm]{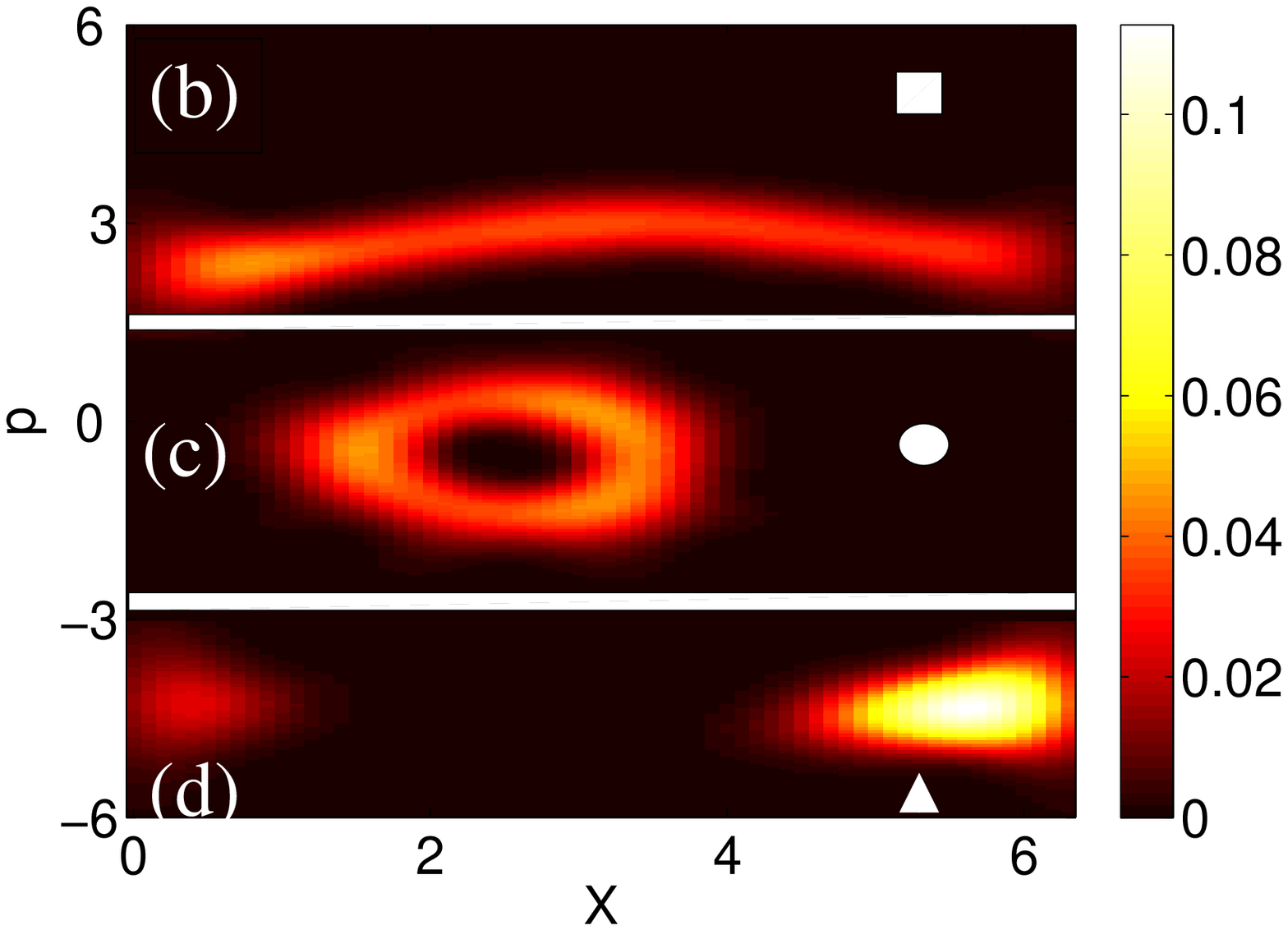}
\end{tabular}
\label{Figure1} \caption{(a) Poincar\'{e} section for the classical
limit; (b-d) Husimi representations for different Floquet
eigenstates for the Hamiltonian (\ref{eq:ham})  with $\hbar=0.2$
(momentum is in units of the recoil momentum, $p_{r}=\hbar k_{L}$,
with $k_{L}=1$). The parameters are $E_{1}=E_{2}=2$, $\omega=2$,
$\theta=-\pi/2$ and $t_{0}=0$. } \label{Fig:husi}
\end{figure}
\end{center}
The two frequency driving,
\begin{equation}
E(t, t_{0})=E_{1}\cos [\omega (t-t_{0})]+E_{2}\cos [2\omega
(t-t_{0})+\theta], \label{eq:driv}
\end{equation}
ensures that  for $E_1,E_{2} \neq 0$ $S_a$ is always violated. In
addition $S_b$ is violated for $\theta \neq 0,\pm \pi$. The
appearance of a nonzero dc-current $J_{ch}=\lim_{t\rightarrow
\infty} 1/t \int_{t_0}^t p(t') dt'$ in this case is due to a
desymmetrization of the chaotic layer structure
(Fig.\ref{Fig:husi}a). It induces a desymmetrization of the events
of directed motion to the right and left \cite{D&F}. Due to
ergodicity inside the layer, the asymptotic current is independent
of the initial time $t_{0}$, for initial conditions located inside
the chaotic layer. With the specific choice of the driving
(\ref{eq:driv}) it follows $J_{ch}(\theta)=-J_{ch}(-\theta)$ and
$J_{ch}(\theta)=-J_{ch}(\theta+\pi)$ \cite{Flach1}. From
perturbation theory it follows $J_{ch}\sim E_{1}^2 E_{2}\sin \theta$
\cite{Flach1, Flach2}. An efficient sum rule allows to compute the
average current $J_{ch}$ by proper integration over the chaotic
layer \cite{Ketz}.

\subsection{Quantum case}

If the Hamiltonian is invariant under the shift symmetry $S_{a}$
(\ref{eq:Sa}), then the Floquet operator possesses the property, see
Eq.(179) in Ref.\cite{Grif}:

\begin{equation}\label{eq:quantum-symme1}
 U(T,t_{0})
=U^{\maltese}\left(T/2,t_{0}\right)^{T}U
\left(T/2,t_{0}\right).
\end{equation}

Here $U^{\maltese}$ performs a transposition along the codiagonal of $U$.
With (\ref{eq:driv}) $S_a$ is always violated.

\begin{figure}
\includegraphics[width=7cm,height=7cm]{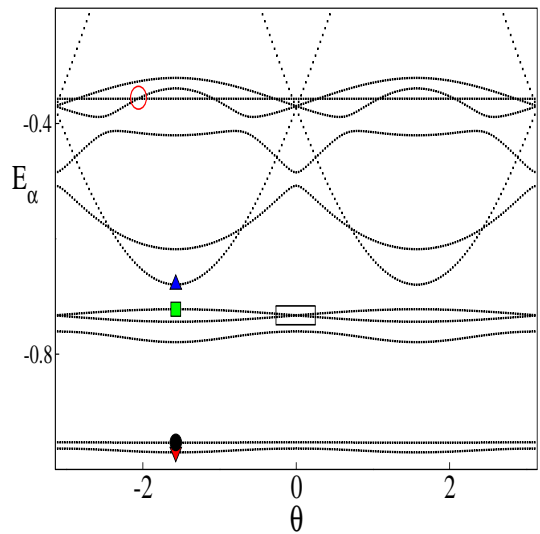}
\caption{ A part of the quasienergy  spectrum  as a function of the
parameter $\theta$. The symbols indicate the corresponding Floquet
states shown in Fig.\ref{Fig:husi}(b-d). The empty red circle
indicates the existence of an avoided crossing between two
eigenstates.
 The empty square
enclose the region of spectrum that appears enlarged in Fig.\ref{Fig:quasidegene-nonquasidene}.}

\label{Fig:spectrum}
\end{figure}

Likewise, one can show that, if the Hamiltonian is invariant under
the time reversal symmetry $S_{b}$ (\ref{eq:Sb}), then the Floquet
matrix has the property \cite{Graham}

\begin{equation}
U(T,t_{0})=U(T,t_{0})^{\maltese}
\end{equation}

 That symmetry will be recovered for $\theta=0,\pi$.
 More details about
the derivation of the previous properties of the operator evolution
are explained in the Appendix \ref{symme}.

We introduce the quantum asymptotic
current as the mean momentum expectation value
\begin{equation}
J(t_0)= \langle \bar{p} \rangle_{t}=\lim_{t\rightarrow \infty} 1/t
\int_{t_0}^{t} dt'\, \langle \psi(t',t_0)|\hat{p}|\psi(t',t_0)\rangle,
\label{eq:cur_def}
\end{equation}

where $\bar{p}=\langle \psi | \hat{p} | \psi \rangle$ is the mean
instantaneous momentum. Expanding the wave function in the Floquet basis the
current becomes
\begin{equation}
J(t_0)=\sum_{\alpha} \langle p \rangle_{\alpha}
|C_{\alpha}(t_{0})|^{2}, \label{eq:current}
\end{equation}
where $\langle p \rangle_{\alpha}$ is the mean momentum of the
Floquet state $|\phi_{\alpha}\rangle$ (see Appendix
\ref{currentformula} for more details). Thus,  we have to study  the
 properties of Floquet states.

In general, the Floquet bands of the system (\ref{eq:Schrodin}-\ref{eq:ham})
 depend on the control parameters like amplitudes, frequencies and phase shifts of the components of
the force $E(t)$. We  vary the parameter
$\theta$ while keeping the other fixed. In Fig.\ref{Fig:spectrum}
we present the quasienergy bands as a function of $\theta$.
There are two
remarkable features.
 First, the spectrum exhibits
two symmetries, $E_{\alpha}(\theta)=E_{\alpha}(-\theta)$ and
$E_{\alpha}(\theta)=E_{\alpha}(\theta +\pi)$, which are consequences
of the choice (\ref{eq:driv}). Second, while some bands show strong
dispersion upon the variation of $\theta$, reaching the maximal
dispersion at $\theta=\pm \pi/2$, others have a rather flat
dependence. In Fig.\ref{Fig:husi}(b-d) we present Husimi functions
\cite{Husimi} for several Floquet states depicted by symbols in
Fig.\ref{Fig:spectrum}. The Planck constant $\hbar=0.2$ is in a
range, where it is possible to establish a correspondence between
different Floquet states and the invariant manifolds of the mixed
phase space for the classical limit. The states (b-d) are located in
various regular phase space regions.

 On the other hand, from Fig.\ref{Fig:spectrum} we observe repulsion between
 some bands whereas other apparently cross each others. These crossings
are in fact avoided crossings which lead to interesting effects.

\begin{figure}
 \begin{center}
\includegraphics[width=8cm,height=7.5cm]{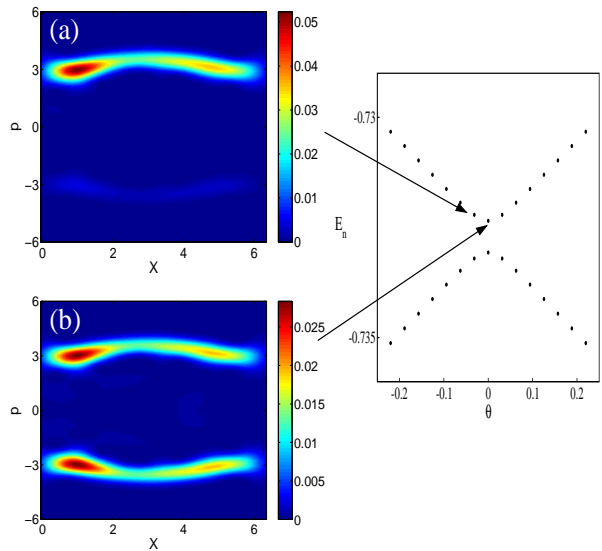}
\caption{Right panel: Quasienergy spectrum in the vicinity of
  two quasidenerated Floquet states. The
symmetry is restored at the points $\theta=0,\pm \pi$ where the
splitting is rather small. Left panels: Husimi function for
different $\theta$ a) $\theta=-0.05$; b) $\theta=0$.}
\label{Fig:quasidegene-nonquasidene}
\end{center}
\end{figure}

For the symmetric case $\theta=0,\pm\pi$ the Floquet matrix has an
irreducible representation using even and odd basis states
$|n\rangle_{s,a} = (|n\rangle \pm|-n\rangle)/\sqrt{2-\delta_{n,0}}$.
All the Floquet states appear as quasidegenerated doublets and
$\langle p \rangle_{\alpha}=0$ for all $\alpha$ (see Figs.\ref{Fig:spectrum}-\ref{Fig:quasidegene-nonquasidene}). We
especially note that this is true for states with arbitrarily large
kinetic energy, for which the corresponding quasienergies become
almost pairwise degenerated. Consequently,  following the
Eq.(\ref{eq:current}), $J=0$ in this case.

For $\theta\neq 0,\pm \pi$ the Floquet states become asymmetric and
the quasidegeneracies are removed
(Fig.\ref{Fig:quasidegene-nonquasidene}). Floquet states acquire a non-zero
mean momentum, thus becoming
transporting. On the other hand, while in the symmetric case it is
possible to have coherent tunneling oscillations between
disconnected regular islands, in the non-symmetric case a possible
dynamical tunneling is suppressed \cite{peres}. Thus, a wave packet
 initially located
on a Floquet state with an asymmetric distribution of momentum (similar
to the state shown in Fig.\ref{Fig:quasidegene-nonquasidene}b) will
undergo a permanent directional transport.

This phenomenon of desymmetrization results in a non trivial
dependence of the momentum upon $\theta$. We can gain some
understanding of this effect by modeling the evolution of two
eigenstates, which form a parity-related pair at the point
$\theta=0$, by neglecting their coupling with the rest of the states.

To this end, we take two eigenstates, $\Phi_{a}$ and $\Phi_{b}$,
which can  be expanded in the plane wave basis $\pm |n\rangle$. The
corresponding propagator for such a reduced system is
\begin{equation}
U_{ab}=\left(%
\begin{array}{cc}
  1-\varepsilon & \bigtriangleup \\
  \bigtriangleup & 1+\varepsilon \\
\end{array}%
\right),
 \label{eq:evolution}
\end{equation}
where the nondiagonal term $\bigtriangleup$ refers to the interaction
between the states. In general, this quantity depends upon the momentum
and driving field's characteristics, i.e.
$\bigtriangleup=\bigtriangleup (p, E_{1}, E_{2}, \theta)$. On the
other hand, the parameter $\varepsilon$ breaks the parity symmetry,
and yields $\varepsilon \sim n E_{1}^{2} E_{2} \sin \theta$ at a
first-order of a perturbation theory. It is known that, in the absence of driving, $E(t)=0$, the asymptotic dependence of the
 splitting value
on $n$ is $\bigtriangleup \sim n^{-n}$, i.e. a superexponential
decay \cite{dorignac}.

The eigenstates  in a plane-wave basis are
\begin{eqnarray}
\Phi_{s}=\frac{1}{L(\gamma)}\left[ 1\cdot |n \rangle + \left(\gamma
+ \sqrt{1+\gamma^{2}}\right) \cdot |-n \rangle \right], \\
\Phi_{a}=\frac{1}{L(\gamma)}
\left[  \left(\gamma
+ \sqrt{1+\gamma^{2}}\right) \cdot |n \rangle -1 \cdot |-n \rangle \right],
 \label{eq:evolution}
\end{eqnarray}
where $\gamma=\varepsilon / \bigtriangleup$ and
$L(\gamma)=\sqrt{1+\left(\gamma+\sqrt{1+\gamma^{2}}\right)^2}$. The
symmetric case, $\theta=0$, corresponds to the limit  $\gamma
\rightarrow 0$. The  eigenstate's momenta are
\begin{eqnarray}
p_{s}\approx -\gamma n, ~~~~ p_{a}\approx  \gamma n,
 \label{eq:evolution}
\end{eqnarray}
At the opposite limit, $\gamma \rightarrow \infty$, we have
\begin{eqnarray}
p_{s}\approx -n\left(1+\frac{1}{4 \gamma^{2}}\right),~~~~
p_{a}\approx  n \left(1+\frac{1}{4 \gamma^{2}}\right).
 \label{eq:evolution}
\end{eqnarray}

For large $n$, $\Delta$ is small, but for $\epsilon=0$
(symmetric case) it follows $\gamma=0$, and the eigenstates carry zero
momentum. Deviating from $\theta=0$, the moments quickly reach values $\pm
n$. That happens for $E_{1}^{2} E_{2} \sin \theta \gg n^{-n-1}$, which will hold for any nonzero $E_{1}$, $E_{2}$, $\theta$, provided $n$ is large enough.

Leaving the region corresponding to the chaotic layer, the splitting
between the doublets drops quite suddenly to zero.
This explains the momentum asymmetry in
Fig.\ref{Fig:quasidegene-nonquasidene} for values  of $\theta$
slightly deviated from $\theta=0$.

\section{Average current and resonances in current rectification}

\begin{center}
\begin{figure}[t]
\begin{tabular}{cc}
\includegraphics[width=6cm,height=6cm]{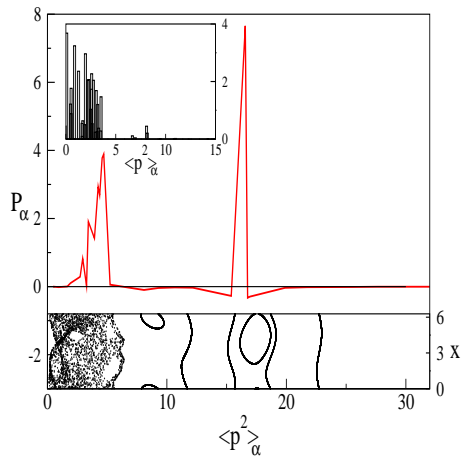}
\end{tabular}
 \caption{Top: The line corresponds to the cumulative momentum,
$P_{\alpha}$ (see text).  All the momenta are scaled in units of the
recoil momentum $p_{r}$. Inset: The bars depict the weights of the
respective Floquet states that overlap with the initial state
$|0\rangle$. The bars have been multiplied by a factor of 10.
Bottom: The plot of the average kinetic energy, complementary to the
Poincar\'{e} section on Fig.1(a). Starting from an initial point
 $(x(t_{0}),p(t_{0})$, we calculate the kinetic energy averaged over
one period, $E_{kin}(mT)=\frac{1}{T}\int_{t_{0}+(m-1)T}^{t_{0}+mT}
p^{2}(t)dt$, and consequently plot points $(x(mT), E_{kin}(mT))$.
The parameters are the same as in Fig.\ref{Fig:husi}
}\label{Fig:weight}
\end{figure}
\end{center}

The  asymptotic current (\ref{eq:cur_def}) depends on the initial
conditions, and, following Eq.(\ref{eq:current}), only those Floquet
states which overlap with the initial wave function, $\psi(t_{0})$,
contribute to its subsequent time evolution. From now on, we
restrict our analysis to the initial condition in the form of a
plane wave with wavevector $n=0$, i.e. $ \psi(t_{0})=|0\rangle
=\frac{1}{\sqrt{2 \pi}}$. This initial condition spreads over
Floquet states with low kinetic energies, namely over those
eigenstates for which the Husimi representations lie in the chaotic
layer's region (Fig.\ref{Fig:weight} ).


In the previous section we have discussed the mechanism of
the dessymetrization when we tuned the system away from the symmetry
point $\theta=0$. In order to estimate a dc-current value we need a
qualitative information about mean momentum values acquired by
Floquet states. As a suitable quantity to study this issue, we use
the cumulative average momentum, $P_{\alpha}=\sum_{s \leq \alpha}
\langle p \rangle_{\alpha}$. Fig.\ref{Fig:weight} depicts the
cumulative average momentum, obtained from the recursive relation
$P_{\alpha+2}=P_{\alpha}+\langle p \rangle_{\alpha+1} +\langle p
\rangle_{\alpha+2}$, with $P_{0}=\langle p\rangle _0$
\cite{recurse}. The asymmetry stems mainly from Floquet states
located in the chaotic layer region of the classical phase space.
The dependence has several peaks. The Floquet states with higher
values of the mean momentum, which yields strong dc-currents, are
located at kinetic energy's regions that correspond to classical
transporting resonance islands (see bottom part of the
Fig.\ref{Fig:weight}). A more detailed analysis by using the Husimi
distribution confirms this finding. There are infinitely many
high-order resonance islands in the phase space of the Hamiltonian
system (2) \cite{Ham_cl}, whose sizes shrink quickly as the
resonance's order increases with $\langle p^{2}\rangle \rightarrow
\infty$. The desymmetrization effect is observable only when a
resonance island can host more than one Floquet state. With
increasing $\langle p^2 \rangle_{\alpha}$, $P_{\alpha}$ approaches
zero in full accordance with the fact that total current over the
whole momentum space should be zero \cite{Ketz}.

\begin{center}
\begin{figure}[t]
\begin{tabular}{cc}
\includegraphics[width=6cm,height=6cm]{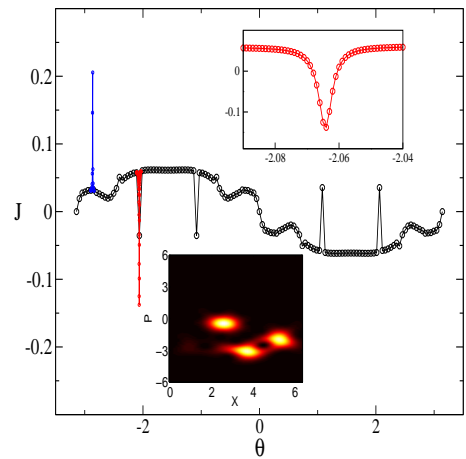}
\end{tabular}
 \caption{ The average current $J$ (in units of the recoil momentum)
vs $\theta$. Blue and red delta points are obtained with higher resolution in
particular $\theta$-regions. Insets:  (top right): Enlargement of the region around
 $\theta=-2.06$. (bottom): Husimi function for the  eigenstates in the avoided
crossing indicated in Fig.\ref{Fig:spectrum}. }
\label{Fig:current0}
\end{figure}
\end{center}

The asymptotic current for a fixed initial condition
$|\psi(t_0)\rangle $ depends in general on the initial time $t_{0}$.
Note that this is possible also in the classical case, since the
initial distribution may overlap with different regular transporting
manifolds \cite{Ketz}. However, if we start with a cloud of
particles  located exactly inside the chaotic layer, the classical
asymptotic current will be independent of $t_0$ for any choice of
the distribution function over the chaotic manifold. This is not
true for the quantum case where the current may even change its sign
with the variation of $t_{0}$. It is a consequence of the linear
character of the Schr\"{o}dinger equation \cite{Qchaos}. We will
first discuss the results obtained after averaging over the initial
time $t_{0}$. Then we can assign a unique current value, $J=1/T
\int_{0}^{T}J(t_{0})dt_{0}$, for fixed parameters of the ac-field,
$E_{1}$, $E_{2}$, and $\theta$. Fig.\ref{Fig:current0} shows the
dependence of the average current on the asymmetry parameter
$\theta$ for the initial condition $|\psi\rangle =|0\rangle $. The
average current $J$ shows the expected symmetry properties
$J(\theta) = -J(\theta+\pi) = -J(-\theta)$. These symmetries for the
current imply that the results obtained for the interval
$(-\pi,-\pi/2)$ holds for other intervals as well.

The typical dependence of the average current $J$ on $\theta$ is
shown on Fig.\ref{Fig:current0}.The curve presents a smooth profile
with several peaks.

By comparing Fig.\ref{Fig:current0}  with the quasienergy spectrum
Fig.\ref{Fig:spectrum}, one can associate peaks with single sharp
avoided crossings (resonances) between two Floquet eigenstates.

 The Husimi function gives us additional insights for
 the appearance of these peaks (see insets in Fig.5). It shows
two states that are located in the chaotic layer, one of them near $p=0$, while
the other is off the line $p=0$ with a strong asymmetry in momentum,
 corresponding to a regular transporting manifold. Off resonance the initial zero-plane wave state
mainly overlaps with the first eigenstate, which
yields some nonzero current due to desymmetrization. At resonance
 this eigenstate mixes with the "transporting" one,  resulting in a strong enhancement
of the current which is reflected in the presence of peaks.

\begin{figure}
\includegraphics[width=5cm,height=8cm]{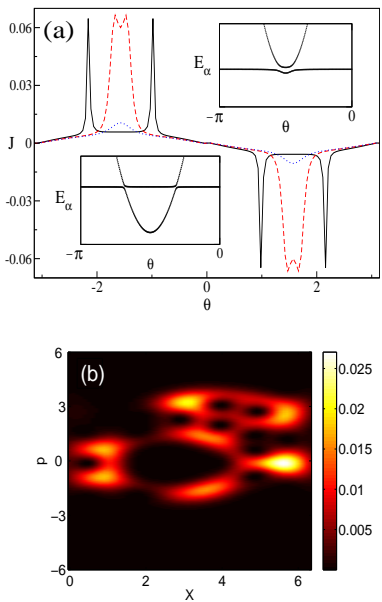}
\caption{ (a) The average current $J$ (in units of the recoil momentum)
vs $\theta$ for different amplitude values of the second harmonic,
$E_{2}$: $0.95$ (pointed line), $1$ (dashed line) and $1.2$ (solid
line). Insets: relevant details of the quasienergy spectrum
versus $\theta$ in the resonance region for
$E_{2}=1$ (top right) and $E_{2}=1.2$ (bottom left).
The parameters are $E_{1}=3.26$ and
$\omega=3$.
(b) Husimi function for the upper eigenstate that appears in (a) (top right inset) with $\theta=-\pi/2$.} \label{Fig:current}
\end{figure}

From an experimental point of view a too narrow resonance may become
undetectable due to resolution limitations. We thus studied how to
vary the width of the resonance without much affecting its
amplitude. It turns out to be possible by tuning another control
parameter, e.g. the amplitude $E_2$. We decrease this field
amplitude in order to disentangle the two Floquet states and remove
the avoided crossing. That will happen for some value of $E_2$ at
$\theta=\pm \pi/2$. The details of the quasienergy spectrum around
that critical point are shown in the insets in
Fig.\ref{Fig:current}a. The two quasienergy spectra disentangle for
$E_2=1$ but stay close over a sufficient broader range of $\theta$
values. Thus the resonances become broader, as seen in
Fig.\ref{Fig:current}a. Further decrease of $E_2$ to a value of 0.95
leads to a strong separation of the two spectra, and consequently to
a fast decay of the amplitude of the resonance.

As we already know, in resonance
Floquet states mix. Here the new eigenstate contain
contributions both from the original chaotic state as well as from
the regular transporting island state (see Fig.\ref{Fig:current}). The Husimi distribution
of this mixed state shows strong asymmetry as expected.
The states from the chaotic sea and the regular transporting island are connected by a
narrow isthmus. Therefore, it is possible to leak from the nonstransporting to
the transporting island, which,
in principle, should reduce significantly the observation time for the
detection of resonances. It was shown in \cite{D&F}, for
the classical limit,
 that for a maximum desymmetrization of the phase space, particles may stick
to a regular island leading to a ballistic
flight. Hence, based on the classical analog, we can say that our hybrid
state serves as a quantum ballistic channel.

\begin{figure}
\includegraphics[width=7cm,height=6cm]{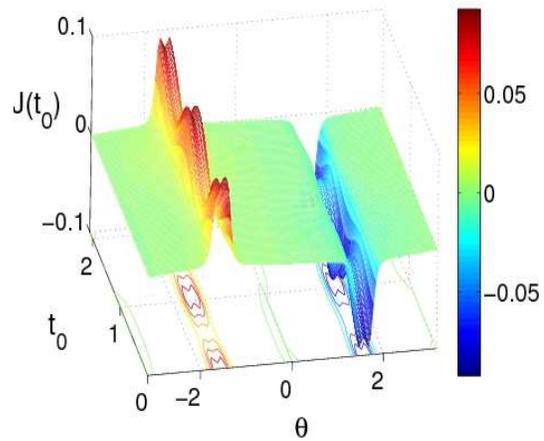}
\caption{Current dependence on the initial time $t_0$ and $\theta$.
The
 parameters are the same as in Fig.\ref{Fig:current}.}
\label{Fig:temp}
\end{figure}

On the other hand, as already mentioned, the asymptotic current
depends on the initial time $t_{0}$. The observed resonance
structures, due to resonant interaction between Floquet states
(avoided crossings of quasienergies), are independent of the initial
time $t_0$. In Fig.\ref{Fig:temp} we plot the nonaveraged current as
a function of both $\theta$ and $t_0$. While the smooth background
is barely resolvable with the naked eye, the resonances are clearly
seen, and their position is not depending on $t_0$, while their
amplitude does. That implies that one can further maximize the
resonant current by choosing properly the initial values for $t_0$.

\begin{center}
\begin{figure}
\includegraphics[width=6cm,height=5.6cm]{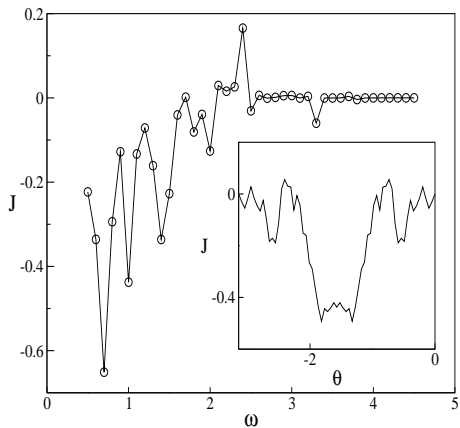}\vspace{0.4cm}
\caption{ The average of current $J$ versus
frequency $\omega$ of the driving force for $\theta=-\pi/2$.  Inset: The average of current $J$ versus
$\theta$ for $\omega=1$. The plot of the
current is depicted for the interval
$(-\pi/2,0)$, since the current is antisymmetric with respect to a
reflection at the origin.  The
other parameters are $E_{1}=3$, $E_{2}=1.5$. }\label{Fig:resonan}
\end{figure}
\end{center}

So far, we have considered only the frequencies, $\omega=2,3$
(Figs.1-7).
To gain a better understanding of the resonances, we compute
the dependence of the current as a function of the frequency of the
driving force (see Fig.\ref{Fig:resonan}). This figure shows the appearance of
peaks, whose amplitudes increase as the frequency decreases.
However, for larger frequencies, peaks show up for specific frequencies, a
result which depends on the drive amplitudes.
Interestingly, one can notice the inversion of current.

In order to increase the current, we reduce the frequency of
 the ac-force $E(t)$.
The asymptotic current for $\omega=1$ is shown in the inset
of Fig.\ref{Fig:resonan}.
We again observe a clear broad resonance, but the maximum current
value increases by an order of magnitude up to $0.5$ in units of
recoil momenta.

Real experiments are always limited by a finite observation time. Therefore,
the rate of convergence to the asymptotic current value, Eq.(12),
becomes a crucial issue. To give a more realistic approach
of our above results to experiments, we perform numerical computations using the running average momentum
(current) definition,
\begin{equation}
 P=\frac{1}{t}\int_{0}^{t} \bar{p} dt.
\end{equation}

\begin{figure}
 \begin{center}
\includegraphics[width=7cm,height=6cm]{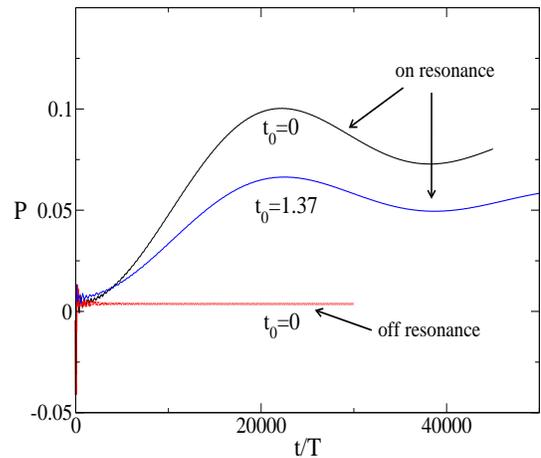}\vspace{0.4cm}
\includegraphics[width=7.5cm,height=6cm]{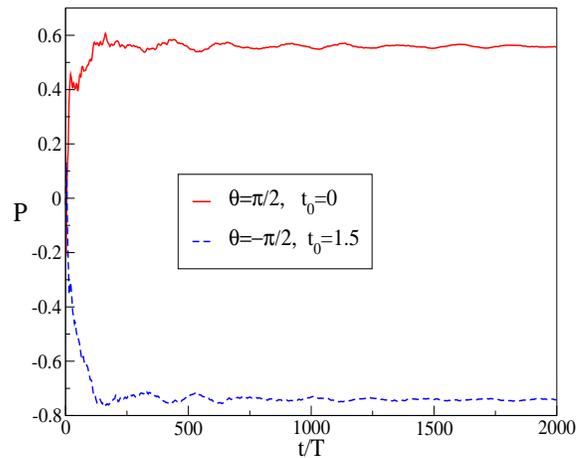}
\caption{Momentum evolution of state $|0 \rangle$ in units of periods. Top
  panel:  On resonance cases $\theta=-\pi/2$ for different $t_{0}$. Off resonance case $\theta=-2.2$.
The parameters are $\omega=3$, $E_{1}=3.26$, $E_{2}=1$.
 Bottom panel: Case $\omega=1$ with parameters  $E_{1}=3$, $E_{2}=1.5$.}\label{Fig:evol}


\end{center}
\end{figure}

The time evolution for the instantaneous momentum $P(t)$ is shown in
Fig.\ref{Fig:evol}. Top of Fig.\ref{Fig:evol} shows that off resonance the momentum rapidly evolves towards an asymptotic
value, while in resonance the momentum performs long term oscillations which slowly
approach the asymptotic value as the time increases. To speed up the
convergence of the momentum towards the asymptotic value, we reduce the frequency. Fig.\ref{Fig:evol} shows large currents close to
the recoil momentum with relaxation times of the order of 100 to 200 periods for $\omega=1$.
The above results were obtained for $\hbar=0.2$. We have repeated the
calculations using $\hbar\sim 1$ and observed similar results.

\section{Quantum Hamiltonian ratchets with flashing potentials}

 Recent
experiments have shown the possibility to achieve an optical lattice
with variable asymmetry \cite{martin}.




The Hamiltonian in this case is given by

\begin{equation}\label{eq:hamil-flashing}
H(x,p,t)=\frac{p^{2}}{2}+U(x)E(t-t_{0}),
\end{equation}

where $U(x+L)=U(x)$ and $E(t+T)=E(t)$.

Here, we take, as in \cite{martin}, the potential

\begin{equation}\label{eq:potential}
U(x)=K\left[\cos(x)+s \cos(2x+\theta_{p})\right],
\end{equation}

where $\theta$ is the parameter which makes the potential asymmetric
for values different from $0, \pm \pi$. The relevant symmetries for
the classical limit of the Hamiltonian (\ref{eq:hamil-flashing}) are

\begin{eqnarray}
\mathcal{S}_{1}: (x, p, t) \rightarrow (-x+2x_{s}, -p, t) ~~~ \label{eq:S1} \\
\mbox{if} ~~~ U(x+x_{s})=U(-x+x_{s})~ \mbox{for some}~  x_{s},
\nonumber
\end{eqnarray}

\begin{eqnarray}
\mathcal{S}_{2}: (x, p, t) \rightarrow (x, -p, -t+2t_{s})
~~~\label{eq:S2}
\\ \mbox{if} ~~~ E(t+t_{s})=E(-t+t_{s})~ \mbox{for some}~  t_{s}, \nonumber
\end{eqnarray}

\begin{eqnarray}
\mathcal{S}_{3}: (x, p, t) \rightarrow (-x+2x_{s},-p,t+T/2) ~~~~~~
\label{eq:S3}
\\ \mbox{if}
~~ E(t)=-E(t+T/2)~\mbox{and}~U(x+x_{s})=-U(-x+x_{s}), \nonumber
\end{eqnarray}

\begin{eqnarray}
\mathcal{S}_{4}: (x, p, t) \rightarrow (x+L/2, -p, -t+2t_{s}) ~~~~~~
\label{eq:S4}
\\ \mbox{if}
~ E(t+t_{s})=-E(-t+t_{s})~\mbox{and}~U(x)=-U(x+L/2). \nonumber
\end{eqnarray}

 It is easy to see that a
two-harmonic potential Eq.(\ref{eq:potential}) is insufficient to break the
time-reversal symmetry $\mathcal{S}_2$ (\ref{eq:S2}).  Thus an asymmetric ac-modulated
function $E(t)$ is needed to desymmetrize the system. We  use the two-harmonic
function (\ref{eq:driv}), which ensures that for $\theta_{p}\neq 0,
\pm\pi$ and $\theta\neq 0, \pm\pi$ all the relevant symmetries,
Eqs.(\ref{eq:S1},\ref{eq:S2},\ref{eq:S3},\ref{eq:S4}), are violated.
As stated before the quantum system inherits the same symmetries as
the classical limit. Therefore a current should appear as the symmetries are
 broken.



\begin{figure}[tbp]
\begin{center}
\includegraphics[width=8cm,height=6cm]{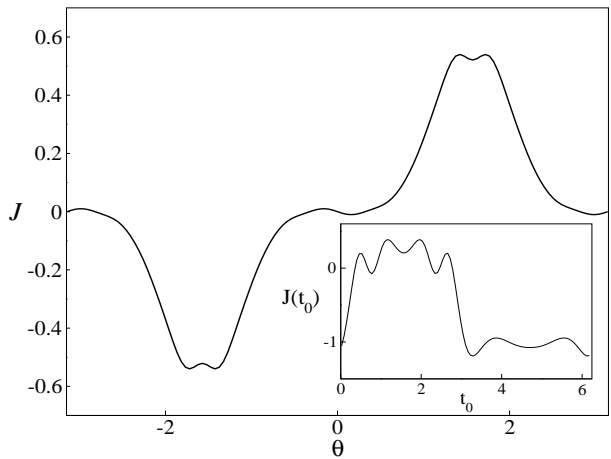}
\end{center}
\caption{The average current $J$ vs $\theta$. Inset:
Asymptotic current as a function of the initial condition $t_{0}$ at
$\theta=-\pi/2$. The parameters are $E_{1}=2$, $E_{2}=1.5$,
$\hbar=1$, $\omega=1$, $K=1.5$, $s=0.25$, $\theta_{pot}=\pi/2$.
}\label{tercera}
\end{figure}

Fig.\ref{tercera} shows the average current $J$ versus $\theta$ for
the asymmetric potential, $\theta_{p}=-\pi/2$, at strong quantum
limit $\hbar=1$. Since the current dependence possesses the symmetry
$J(\theta) = -J(\theta+\pi) = -J(-\theta)$, it is sufficient to plot
the current for the interval $\theta \in (-\pi,0)$. As for the case
of rocking ratchets, we found evidence for quantum resonances and
dependence of the asymptotic current on the initial time $t_{0}$
(see inset in Fig.\ref{tercera}).

\section{Concluding remarks}

In summary, we have investigated the mechanisms of the emergence of
an average current  in ac-driven quantum systems. The key mechanism
of the current rectification  is the desymmetrization of Floquet
states. A peculiarity of the quantum ratchet is its strong
dependence of the current on the chosen initial time $t_{0}$ of the
applied field.  We identified quantum resonances induced by avoided
crossings between Floquet states which enhance the current
drastically. Optimizing the drive frequency, amplitude and initial
phase, resonant currents easily reach the recoil momentum value and
should be experimentally observable using driven cold atoms in
optical lattices \cite{ren1, martin}.\\

\acknowledgments {This work has been supported by the DFG-grant
HA1517/31-1 (P. H. and S. F.).}

\appendix

\section{Gauge transformation: Some relevant quantities}\label{gauge}

We introduce the transformation
\begin{equation}
|\tilde \psi \rangle =\nonumber \exp\left[\frac{i}{\hbar}x
A(t)\right] |\psi \rangle
\end{equation}
with $A(t)=-\int_{0}^{t}E(t')dt'$.

Substituting into Eq.(\ref{eq:Schrodin}) we can rewrite the Schr\"odinger equation as
\begin{equation}
i \hbar \frac{\partial}{\partial t} |\tilde \psi(t)\rangle =
\tilde H(t)|\tilde \psi(t)\rangle, \label{shrogauge}
\end{equation}
where $\tilde H(t)=\frac{1}{2}
\left[\hat{p}-\frac{A(t)}{\hbar}\right]^2 +[1+\cos(x)]$.

To compute the momentum we start from the definition
\begin{eqnarray}
\tilde p(t)=\langle \tilde \psi |-i \hbar \frac{\partial}{\partial
x}|\tilde \psi\rangle =~~~~~~~~~~~ \nonumber
\\ -i \hbar\int^{\infty}_{-\infty}
\tilde \psi^* \frac{\partial}{\partial x} \tilde \psi dx=p(t)+A(t)
\end{eqnarray}

where $\langle  \tilde{p} \rangle=\langle  p \rangle $.

The kinetic energy $\mathcal{T}$ is given by
\begin{eqnarray}
\mathcal{T}(t)=\left\langle \psi \left\vert-\hbar^2 \frac{\partial^2}{\partial x^2} \right\vert \psi \right\rangle=~~~~~~~~~~~
\nonumber \\
\left\langle \exp\left[\frac{i}{\hbar}x A(t)\right] \tilde{\psi}
\left \vert -\hbar^2 \frac{\partial^2}{\partial x^2} \right\vert
 \exp\left[-\frac{i}{\hbar}x A(t)\right] \tilde{\psi} \right\rangle.
\end{eqnarray}

After straightforward calculations
\begin{equation}
\mathcal{T}(t)= \tilde{ \mathcal{T}} (t)- 2A(t)
\tilde{\mathcal{T}}(t) + A(t)^2.
\end{equation}
The average kinetic energy reads
\begin{equation}
\langle \mathcal{T} \rangle=\langle \tilde{\mathcal{T}} \rangle
-2\langle A(t) \tilde{\mathcal{T}}(t)  \rangle +\langle A(t)^2
\rangle.
\end{equation}

Recalling the Husimi
distribution,
\begin{equation}
\rho(\langle  x \rangle,\langle p \rangle ) =\frac{1}{2\pi}|\langle \psi|
\Phi_{\langle  x \rangle,\langle p \rangle}  \rangle |^2,
\end{equation}
 we can recast it in the new frame of variables as,

\begin{eqnarray}
\tilde \rho(\langle  x \rangle,\langle p \rangle)&&=
\frac{1}{2\pi}\left \vert  \int^{\infty}_{-\infty} e^{-i
\frac{x}{\hbar} A(t)}\psi^*(x,t) e^{i\langle p
  \rangle x + \frac{(\langle x \rangle-x)^2}{4\sigma^2}}  dx \right \vert^2 \nonumber \\
&&\frac{1}{2\pi} \left \vert \int^{\infty}_{-\infty} \psi^*(x,t)  e^{i[\langle p
  \rangle -\frac{A(t)}{\hbar}]x + \frac{(\langle x \rangle-x)^2}{4\sigma^2}}  dx\right \vert^2
\nonumber \\
&&=\rho\left\{\langle  x \rangle,\langle p \rangle
-\frac{A(t)}{\hbar}\right\}.
\end{eqnarray}
Finally, after reversing the operations, we find

\begin{equation}
 \rho(\langle  x \rangle,\langle p \rangle)=\tilde \rho\left\{\langle  x
 \rangle,\langle p \rangle +\frac{A(t)}{\hbar}\right\}.
\end{equation}

\section{Numerical integration schemes}\label{nume}

In order to find the numerical solution for the time-dependent Schr\"odinger
equation we have used two different methods of integration.
The first method has been described in \cite{gauge}.
It starts by using the expansion Eq.(\ref{Eq:wave}) and approximating
\begin{equation}
\hat{A}(t)=\frac{\Delta t}{\hbar}  A(t) \sum_{k}\delta(t-k \Delta t),
\end{equation}
for the time dependence of the vector potential
$A(t)=-\int_{t_{0}}^{t}E(t')dt'$, with $\Delta t=2\pi /(\omega
N_t)$, where $N_t$ is the number of integration steps per period.

The integration is carried out computing the coefficients $c_{n}(t_{k}+\Delta
t)$ ($t_k=k\Delta t$) by successive multiplication of the vector $c_{n}(t_{k})$ by
a unitary matrix
\begin{equation}
c(t_{k}+\Delta t)=UQVQ^{-1}c(t_{k})
\end{equation}
where  $U$ and $V$ are diagonal matrices with
\begin{equation}
U_{n,n}=\exp \left \{ -i\hbar \frac{ \Delta t}{2} \left[ n^2 -2 n \frac{ A(t)}{\hbar}+ \frac{A(t)^2}{\hbar^2}
  \right]\right \}
\end{equation}
and $V_{n,n}=\exp(-i \Delta t \tilde{v}_{n})$.

$\tilde{v}_{n}$ are the  eigenvalues  of the matrix representation $M$ of the
operator $(1/\hbar) \cos(\hat{x})$. Q is an orthogonal (unitary) matrix that
transforms
$M$ into diagonal form ($ M=Q \tilde{ V} Q^{-1}$). $\tilde{ V}$ is a diagonal
matrix with elements $\tilde{v}_{n}$.

The symmetric matrix $M$ has the structure
$M_{n,m}=(\delta_{m,n+1}+\delta_{m,n-1})/(2 \hbar)$. This method
permits to decrease the number of steps per period \cite{casati}.
Nevertheless, it involves several matrix products and therefore its
efficiency is degraded.

 To check our results we use a
second integration method which is more efficient \cite{Graham}.
We expand again over plane waves but this time we write down the expansion
coefficients as a product of two time dependent functions, viz.
\begin{equation}\label{ansatz}
|\psi(t)\rangle=\sum_{n=-N}^{N} a_{n}(t) \phi_{n}(t) |n \rangle.
\end{equation}

Assuming that $\phi_{n}$ is a solution of the problem
\begin{equation}
i\hbar \partial_{t}\phi_{n}(t)=\frac{\hbar^2}{2}\left[ n-\frac{A(t)}{\hbar}
\right]^2 \phi_{n},
\end{equation}
we obtain for $\phi_{n}(t)$ the solution
\begin{eqnarray}
\phi_{n}(t)=\exp \left \{ -i \frac{\hbar}{2} \left[n^2 t -\frac{2n}{\hbar}
  \int_{0}^{t}A(t')dt' \right. \right. \nonumber \\
\left. \left. +\frac{1}{\hbar^2} \int_{0}^{t} A(t')^2 dt'
  \right]\right \}.
\end{eqnarray}
Then inserting Eq.(\ref{ansatz}) into Eq.(\ref{shrogauge}) we obtain the
set of equations for the amplitudes $a_{n}(t)$
\begin{equation}
\dot{a}_{n}=-i\frac{1}{2 \hbar}\left[ \frac{\phi^t_{n+1}}{\phi^t_{n}}a_{n+1}+
 \frac{\phi^t_{n-1}}{\phi^t_{n}}a_{n-1}\right].
\end{equation}

To obtain the propagator $U(T,0)$, we start with the initial states
$a_{n}^k=\delta_{n,k}$, where $-N \leq n,k \leq N$ and integrate over one
time period $T$. Here the column of the matrix for the propagator is given
by $a_{n}(T)\phi_{n}(T)$.
For the computation we have
  neglected the contribution originating from $A^{2}(t)$ and the constant term
  $1/\hbar$ in the potential since they only yield global phase factors.

\section{Symmetries of the Floquet operator}\label{symme}

Let us first analyze the time reversal symmetry:
$$S_{b}: t \rightarrow -t, \qquad x \rightarrow -x .$$
In the momentum representation basis, $|n\rangle$, this
transformation implies several operations including complex conjugation, and
inversions  $n \rightarrow -n $.

In that case the Floquet operator possesses the property
$$U(T,t_{0})\equiv U(T,t_{0})^{\maltese},$$
whose matrix elements obey
$$u_{n,m}=u_{-n,-m}.$$
The latter relation holds whenever the transformation is done around
the symmetric points, namely $t_{0}=t_{s}$.

Let us analyze the symmetry

$$S_{a}: t\rightarrow t+T/2, \qquad x \rightarrow -x.$$
This corresponds to the transformation
$$t\rightarrow  t+T/2,  \qquad n \rightarrow -n.$$

For the Floquet operator we get
\begin{eqnarray}
U(T,0)\equiv U(T,T/2)U(T/2,0) \nonumber \\  ~~~~~~~~~~~~~~~~~~\equiv
U(T/2,0)^{\maltese T}U(T/2,0).
\end{eqnarray}

This implies that
$$U \rightarrow U^{\maltese T},$$ or, equivalently,
$$u_{n,m}=u_{-n,-m}.$$

\section{Asymptotic current}\label{currentformula}

The asymptotic current can be defined as follows
\begin{eqnarray}\label{asimp}
 J=\lim_{\mu \rightarrow \infty} \frac{1}{\mu}
\sum_{m=1}^{\mu} \langle\psi(mT)|\hat{p}|\psi(mT) \rangle_{T} =\nonumber
\\
\lim_{\mu \rightarrow \infty} \frac{1}{\mu}
\sum_{m=1}^{\mu}\langle \langle p(mT) \rangle\rangle_{T}.
\end{eqnarray}

The wave function can be expanded over Floquet states as
\begin{equation}
|\psi(t)\rangle=\sum_{\alpha} C_{\alpha} |\phi_{\alpha} \rangle.
\end{equation}
Similarly, one can expand the Floquet states over the plane wave
function
\begin{equation}
|\phi_{\alpha}(t)\rangle=\sum_{n} b_{\alpha, n} |n \rangle.
\end{equation}

Starting from an initial plane wave $|0\rangle$, the
coefficients become $C_{\alpha}=b_{\alpha, 0}$. In
such a case we get
\begin{equation}
 \langle p(mT) \rangle =\sum_{\alpha,\alpha'} b_{\alpha,0} b^{*}_{\alpha',0} e^{-i m
      (E_{\alpha}-E_{\alpha'})}
\langle
\phi_{\alpha'} |\hat{p} |\phi_{\alpha} \rangle.
\end{equation}

It was shown  for chaotic systems \cite{delande} that, as the time
goes on, nonlinear interference terms accumulate large and larger
phases, namely $m(E_{\alpha}-E_{\alpha'})$. Finally, for long enough times, contributions to the directed
transport are only given by diagonal terms,

\begin{equation}
 J =\langle \sum_{\alpha} |b_{\alpha,0}|^2
\langle
\phi_{\alpha} |\hat{p} |\phi_{\alpha} \rangle \rangle_{T}.
\end{equation}


\begin{thebibliography}{1000}
\bibitem{gauge1} When calculating Husimi distributions, average
kinetic energies, etc., we  used the inverse gauge transformation, in
order to return to the original wave function.
\bibitem{recurse} We introduced a summation by pairs in order to avoid
an "oscillating" behaviour of $P_{\alpha}$ for large values of
$\alpha$ (see Eq.(20)).
\end{thebibliography}
\end{document}